\documentclass{PoS}

\title{2HDM Charged Higgs Boson Searches
at the LHC: Status and Prospects}

\ShortTitle{2HDM $H^\pm$ Searches@LHC}

\author{\speaker{Stefano Moretti}\\
        School of Physics and Astronomy, University of Southampton,\\
	Southampton, SO17 1BJ, United Kingdom\\
        E-mail: \email{s.moretti@soton.ac.uk}}

\abstract{We review status and prospects of searches for the charged Higgs boson of 2-Higgs Doublet Models of all Yukawa types 
at the Large Hadron Collider. }

\FullConference{Prospects for Charged Higgs Discovery at Colliders\\
		3-6 October 2016\\
		Uppsala, Sweden}

\begin{document}

\section{Introduction}

Following the discovery of a neutral Higgs boson (herafter denoted by $h$) at the Large Hadron Collider (LHC) in July 2012   \cite{Aad:2012tfa,Chatrchyan:2012xdj}, the quest for new physics Beyond the Standard Model (BSM) must account for a Electro-Weak Symmetry Breaking (EWSB) dynamics governed by the Higgs mechanism. As the discovery of the $h$ state corresponds to that of the last 1/4 of
a (complex) doublet Higgs field\footnote{In fact, 3/4 of it were discovered at the S$p\bar p$S in the form of the $W^\pm$ and $Z$ bosons.}, it makes sense to investigate BSM scenarios which embed such specific Higgs
fields. With this in mind, it is clear that the simplest BSM realisation of an EWSB scenario based on the Higgs mechanism 
is the one afforded by 2-Higgs Doublet Models (2HDMs) \cite{Branco:2011iw}, wherein two Higgs fields, $\Phi_1$ and $\Phi_2$, are introduced.
On the one hand, these scenarios allow for the existence of a SM-like Higgs state
(alignment limit), in accordance with the experimental findings of the ATLAS and CMS collaborations \cite{Khachatryan:2014jba, Aad:2015ona}. On the
other hand, they offer a variety of new Higgs states potentially accessible at the LHC, i.e., another CP-even field ($H$), a 
CP-odd one ($A$) as well as, most notably, a charged pair ($H^\pm$). 

The production and decay rates of the latter would depend upon specific details of the underlying
2HDM \cite{Gunion:1989we}, especially the Yukawa interactions.
Since such an extended Higgs sector naturally leads to Flavour Changing
Neutral Currents (FCNCs), these would have to be suppressed \cite{Glashow:1976nt,Paige:1977nz}. This is
normally achieved by imposing discrete symmetries in modeling the
Yukawa interactions. 

The purpose of this write-up is to  review status and prospects of searches for 2HDM $H^\pm$ states at the LHC. 
In doing so, we borrow several elements from a recent review touching on the same topic \cite{Akeroyd:2016ymd}.
The plan of this note is as follows: in Sect.~\ref{sec:Lagrangian} we describe the $H^\pm$ interactions within the 2HDMs and list 
theoretical and experimental contraints. Sects.~\ref{sec:present} and \ref{sec:future} cover present and future  $H^\pm$ studies at the CERN collider, respectively. Finally, we conclude in Sect.~\ref{sec:summa}. 

\section{$H^\pm$ Couplings in   2HDMs}
\label{sec:Lagrangian}

We limit ourselves to studying the softly $Z_2$-violating 2HDM potential, which  reads \cite{Akeroyd:2016ymd}
\begin{eqnarray} \label{Eq:fullpot}
V(\Phi_1,\Phi_2)=
& -&\frac12\left\{m_{11}^2\Phi_1^\dagger\Phi_1
+ m_{22}^2\Phi_2^\dagger\Phi_2 + \left[m_{12}^2 \Phi_1^\dagger \Phi_2
+ {\rm{h.c.}}\right]\right\} \nonumber \\
& + &\frac{\lambda_1}{2}(\Phi_1^\dagger\Phi_1)^2
+ \frac{\lambda_2}{2}(\Phi_2^\dagger\Phi_2)^2
+ \lambda_3(\Phi_1^\dagger\Phi_1)(\Phi_2^\dagger\Phi_2)
+ \lambda_4(\Phi_1^\dagger\Phi_2)(\Phi_2^\dagger\Phi_1)
\nonumber \\
& + &\frac12\left[\lambda_5(\Phi_1^\dagger\Phi_2)^2 +{\rm{h.c.}}\right].
\end{eqnarray}
Apart from the term $m_{12}^2$, this potential exhibits a $Z_2$ symmetry,
\begin{equation} \label{Eq:Z2-symmetries}
(\Phi_1,\Phi_2) \leftrightarrow (\Phi_1,-\Phi_2) \quad{\rm{or}} \quad
(\Phi_1,\Phi_2) \leftrightarrow (-\Phi_1,\Phi_2).
\end{equation}
The most general potential contains in addition two more quartic terms, with coefficients $\lambda_6$ and $\lambda_7$,  and violates the $Z_2$ symmetry in a hard way \cite{Gunion:1989we}.
The
parameters $\lambda_1$--$\lambda_4$, $m^2_{11}$ and $m^2_{22}$ are real.  There
are various bases in which this potential can be written, often they are
defined by fixing properties of the vacuum state.  The potential (\ref{Eq:fullpot}) can lead to
CP violation, provided $m_{12}^2\neq 0$. Upon EWSB, of the 8 degrees of freedom of $\Phi_1$ and $\Phi_2$, 3 are absorbed as
scalar polarisations of the $W^\pm$ and $Z$ gauge vectors while the remaining 5 appear as physical Higgs states ($h, H, A$ and $H^\pm$).

\subsection{Gauge Couplings}
With all momenta incoming, we have the $H^\mp$ gauge couplings \cite{Gunion:1989we}:
\begin{eqnarray}
H^\mp W^\pm h&:&\quad
\frac{\mp ig}{2}
\cos(\beta -\alpha)
(p_\mu-p_\mu^\mp), \nonumber \\
H^\mp W^\pm H&:&\quad
\frac{\pm ig}{2}
\sin(\beta -\alpha)
(p_\mu-p_\mu^\mp),  \nonumber \\
H^\mp W^\pm A&:&\quad
\frac{g}{2}
(p_\mu-p_\mu^\mp).
\label{Eq:CPC-gauge-couplings}
\end{eqnarray}
Here, $\tan\beta$ is the ratio of the Vaccum Expectation Values (VEVs) of the 2 doublets $\Phi_1$ and $\Phi_2$, which is typically defined
between 1 and $\sim m_t/m_b$. Further, $\alpha$ is the mixing angle in the CP-even Higgs sector, its range being $\pi$, e.g., $[-\pi/2,\pi/2]$. 
The strict SM-like limit corresponds to $\sin(\beta-\alpha)=1$,
however, the experimental data from the LHC
\cite{Khachatryan:2014jba,Aad:2015ona} allow for departures from it.

\subsection{{Yukawa Couplings}} 
\label{sect:Yukawa}
\setcounter{equation}{0}

There are various ``Types'' of Yukawa interactions, all of them can lead to the
suppression of  FCNCs at the tree-level, assuming some vanishing Yukawa matrices.
The most popular is Type-II,
in which up-type quarks couple to one ($\Phi_2$) while
down-type quarks and charged leptons couple to the other scalar
doublet ($\Phi_1$). 
They are presented schematically in Tab.~\ref{tab:Z2}, wherein the symbols $u$, $d$ and $\ell$ refer to up-, down-type quarks and charged leptons of any generation, respectively.

\begin{table}
\centering
\begin{tabular}{cccc}
\hline
Model & $d$ & $u$ & $\ell$ \\
\hline
I & $\Phi_2$ & $\Phi_2$ & $\Phi_2$ \\[0.1cm]
II & $\Phi_1$ & $\Phi_2$ & $\Phi_1$ \\[0.1cm]
X & $\Phi_2$ & $\Phi_2$ & $\Phi_1$ \\[0.1cm]
Y & $\Phi_1$ & $\Phi_2$ & $\Phi_2$ \\[0.1cm]
\hline
\end{tabular}
\caption{The most popular  Yukawa interactions for 2HDMs. 
  Here, $\Phi_1$ and $\Phi_2$ refer to the Higgs doublet coupled to the particular fermion.} 
\label{tab:Z2}
\end{table}

Explicitly, for the charged Higgs boson in Type-II, we have for the
coupling to, e.g.,  the third generation of quarks \cite{Gunion:1989we}:
\begin{eqnarray}  \label{Eq:Yukawa-charged-II}
H^+ b \bar t&: &\quad
\frac{ig}{2\sqrt2 \,m_W}\,V_{tb}
[m_b(1+\gamma_5)\tan\beta+m_t(1-\gamma_5)\cot\beta], \nonumber \\
H^-  t\bar b&: &\quad
\frac{ig}{2\sqrt2 \,m_W}\,V_{tb}^*
[m_b(1-\gamma_5)\tan\beta+m_t(1+\gamma_5)\cot\beta].
\end{eqnarray}
For other Yukawa models the factors $\tan\beta$ and $\cot\beta$ are substituted according to Tab.~\ref{tab:couplings}.

\begin{table}[htb]
\begin{center}
\begin{tabular}{| c | c | c | c |} \hline 
&\multicolumn{1}{c|}{$d$}& \multicolumn{1}{c|}{$u$} 
& \multicolumn{1}{c|}{$\ell$}   \\ \hline  \hline
{I} &  $-\cot\beta$ & $+\cot\beta$  & $-\cot\beta$   \\
{II} &  $+\tan\beta$ & $+\cot\beta$  & $+\tan\beta$   \\
{X}  & $-\cot\beta$ & $+\cot\beta$ & $+\tan\beta$  \\ 
{Y} & $+\tan\beta$ & $+\cot\beta$  & $-\cot\beta$  \\ 
\hline
\end{tabular}
\caption{ Yukawa couplings for 2HDMs without tree-level FCNCs normalised to the  SM vertices.}
\label{tab:couplings}
\end{center}
\vskip -0.2cm
\end{table}

\subsection{Theoretical Constraints} 

The 2HDM is subject to various theoretical constraints. First, it has to have a stable VEV  \cite{Nie:1998yn,Ferreira:2004yd,Goudelis:2013uca,Swiezewska:2015paa,Khan:2015ipa}, which leads to so-called positivity constraints for the potential \cite{Nie:1998yn,Deshpande:1977rw,Kanemura:1999xf}, $V(\Phi_1,\Phi_2)>0$ as $|\Phi_1|, |\Phi_2| \to\infty$.
Second, we should be sure to deal with a particular vacuum (a global minimum) as in some cases various minima can coexist \cite{Barroso:2013awa,Ginzburg:2010wa,Swiezewska:2012ej}. 

Other types of constraints arise from requiring tree-level unitarity and perturbativity of the Yukawa couplings
\cite{Kanemura:1993hm,Akeroyd:2000wc,Arhrib:2000is,Ginzburg:2003fe,Ginzburg:2005dt}. 
In general, these constraints limit the absolute values of the $\lambda$ parameters as
well as $M_{H^\pm}$ (which should not be beyond $\approx 700$ GeV) and $\tan \beta$ (both at very low and very high values). This limit is particularly strong for a $Z_2$ symmetric model \cite{Swiezewska:2012ej,WahabElKaffas:2007xd,Gorczyca:2011he}.

\subsection{Experimental Constraints} 

The EW  precision data, parametrised in terms of the so-called $S,T$ and $U$ parameters \cite{Kennedy:1988sn,Peskin:1990zt,Altarelli:1990zd,Peskin:1991sw,Altarelli:1991fk,Grimus:2007if,Grimus:2008nb},   
provide important constraints on 2HDMs  \cite{Agashe:2014kda}. Furthermore, the muon magnetic moment 
\cite{WahabElKaffas:2007xd,Cheung:2003pw,Chang:2000ii,Krawczyk:2002df}
and the electric dipole moment of the electron
\cite{Regan:2002ta,Pilaftsis:2002fe}
 limit  the
charged Higgs sector of 2HDMs. However,  $B$-physics constraints are the strongest ones emerging from low-energy observables. The key ones include 
{{$B\to \tau \nu_\tau (X)$}}, 
{{$B\to D\tau \nu_\tau$}},
{{$D_s\to \tau \nu_\tau$}},
{{$B\to X_s \gamma$}},
{{$B_0-\bar{B}_0$} mixing}.

The ratio $R^0_b \equiv \Gamma_{Z\to b\bar b} /\Gamma_{Z\to {\rm had}}$ would also be affected by Higgs exchange and, while the
contributions from neutral Higgs
bosons are negligible, those from charged ones are sizable  \cite{Denner:1991ie}. Indeed, LEP and  Tevatron have given limits on the $H^\pm$ mass and couplings,
for charged Higgs bosons in  2HDMs. At LEP a lower mass limit of 80 GeV that refers to the Type-II scenario for ${\rm{BR}}(H^+ \to\tau^+ \nu)+{\rm{BR}}(H^+ \to  c \bar s)=100\%$  was derived.  The mass limit for ${\rm{BR}}(H^+\to \tau^+ \nu) = 100\%$  is 94~GeV (95\% Confidence Level (CL)) while for ${\rm{BR}}(H^+\to c \bar s) = 100\%$ the regions below 80.5 GeV and within 83--88~GeV are excluded (95\% CL). 
Searches for the decay mode $H^\pm \to W^\pm A$ with $A\to b\bar b$, which is not negligible in Type-I, leads  to the corresponding $M_{H^\pm}$ limit of 72.5 GeV (95\% CL) if $M_A > 12~{\rm{GeV}}$ \cite{Abbiendi:2013hk}.

A summary of the discussed constraints (e.g., for the 2HDM-II)   performed by the ``Gfitter'' group \cite{Flacher:2008zq} is presented in 
Fig.~1.
The strongest limit comes from $B\to X_s \gamma$ and the recent inclusion of higher-order effects push the $M_{H^\pm}$ constraint up to around 480~GeV \cite{Misiak:2015xwa} (see also Ref.~\cite{Enomoto:2015wbn}). 
\begin{figure}[!t]
\label{Fig:gfitter}
\begin{center}
\includegraphics[scale=0.6]{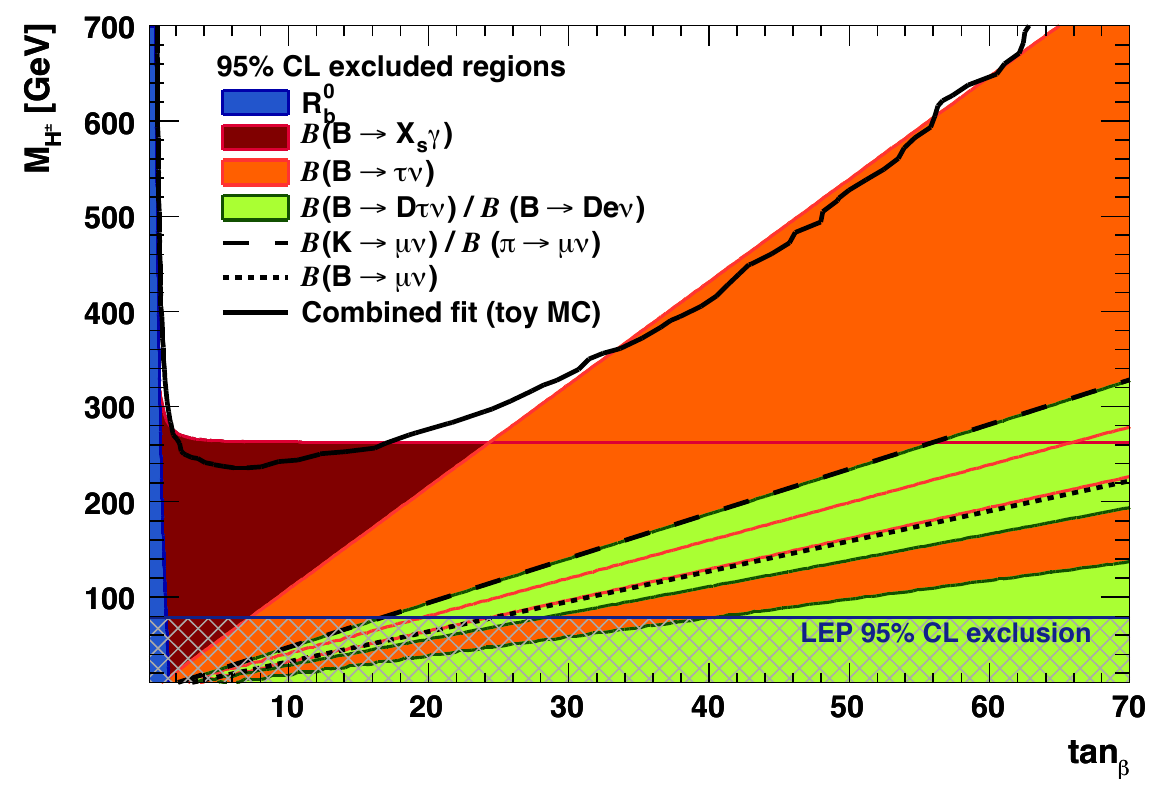}
\end{center}
\vspace*{-8mm}
\caption{Exclusion regions of the 2HDM-II over the [$\tan\beta$, $M_{H^\pm}$] plane at 95\% CL. [Fig. 14 from  \cite{Flacher:2008zq}.]}
\end{figure}


\section{Current LHC Status}
\label{sec:present}

Fig.~2
shows the typical BRs of the $H^\pm$ state in the standard 2HDMs in the case of a light ($M_{H^\pm}<m_t$) and heavy ($M_{H^\pm}>m_t$)
state, for  two representative masses. From these plots, it is clear that in the former mass region the $\tau\nu$ decay is the best one to pursue, given its cleanliness (e.g., in comparison to $cs$)  in the highly QCD-polluted environment of the LHC and its relatively high rates, though also note the role of $cb$ in Type-Y. In the latter mass interval,  it would appear that $tb$ and/or $W^\pm h/H$  can play a significant role (again, alongside $\tau\nu$, which remains relevant in the Type-II and X). In fact, both $tb$ and $W^\pm h/H$ lead to the same signature, $W^\pm b\bar b$, as $t\to bW^+$ and
$h/H\to b\bar b$, so that it is indeed this inclusive mode that ought to be maximised to improve searches in the heavy $M_{H^\pm}$ region \cite{Moretti:2016jkp}, which are notoriously difficult because of the QCD noise. Notice that, in the plot, $M_{H^\pm}=M_A$, so that $H^\pm\to W^\pm A$ decays are forbidden. However, one could swap $H\leftrightarrow A$ and obtain a similar decay pattern. Indeed, this decay (for a very light $A$ state, which is possible unlike the corresponding $H$ case) can play a key role at the LHC Run 2 in a Type-I 2HDM (as we shall see later). Concerning $H^\pm$ production dynamics, this is dominated by the subprocesses $gg,q\bar q\to b\bar b H^+W^-$ ($gg$ largely dominating over $q\bar q $ at the LHC), see
Fig.~3
These contain both $t\bar t$ production and decay (relevant for $M_{H^\pm}<m_t$, 
Fig.~3a)
as well $H^\pm$ Higgs-trahlung (relevant for $M_{H^\pm}>m_t$, 
Fig.~3b)
topologies.

\begin{figure}[htb]
\label{Fig:cpc-br-ratios-vs-mass-3-30}
\begin{center}
\vspace*{-2mm}
\includegraphics[angle=0,scale=0.50]{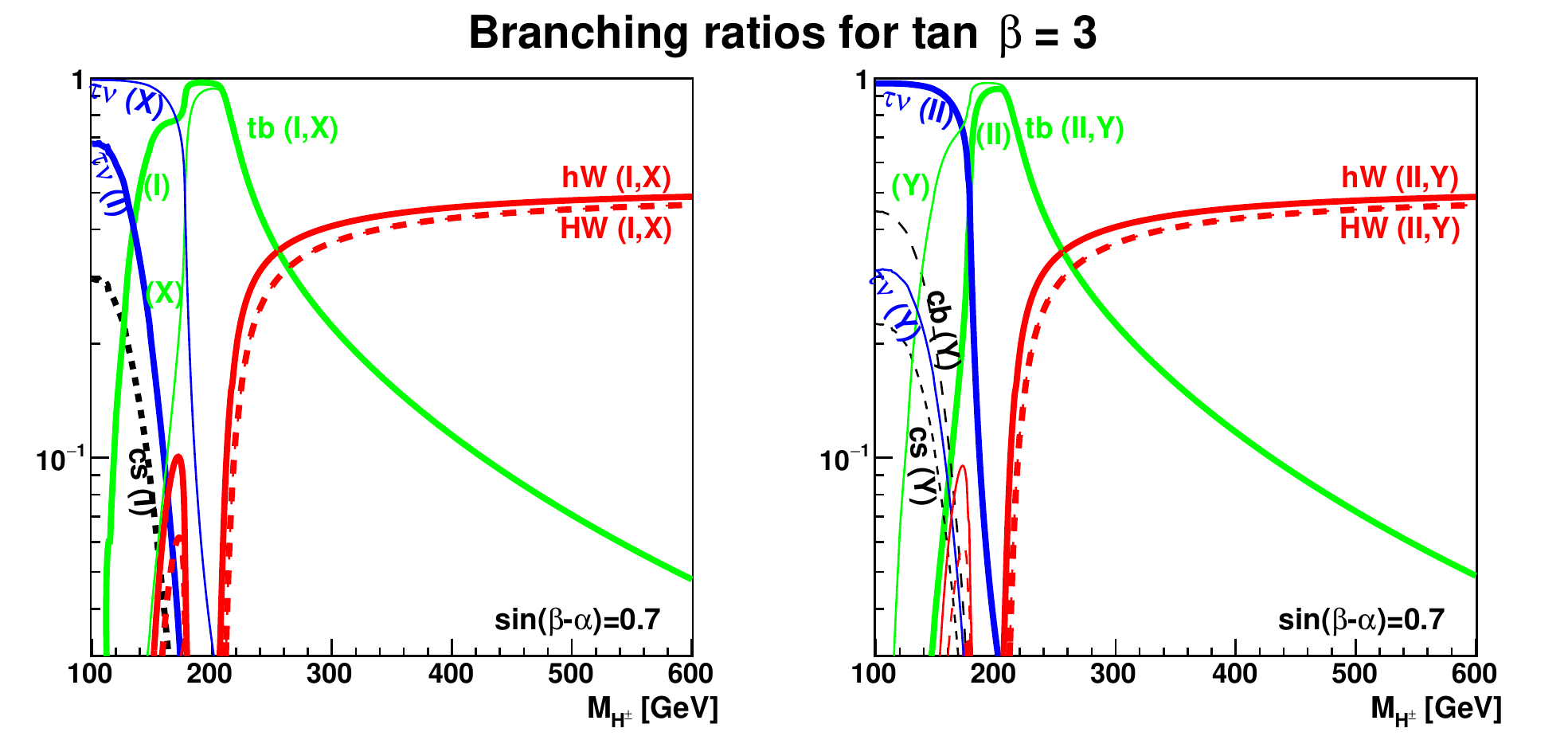}
\includegraphics[angle=0,scale=0.50]{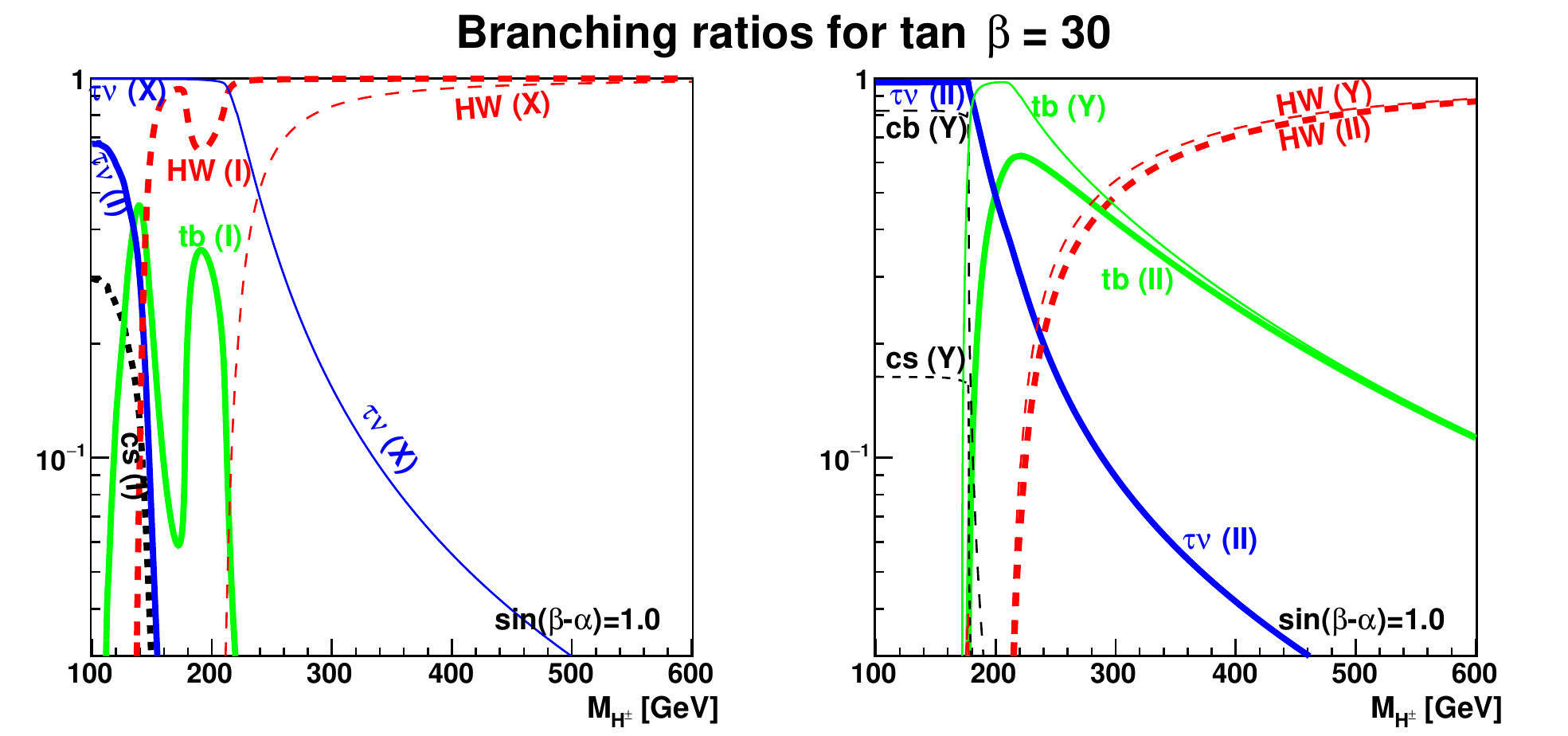}
\end{center}
\vspace*{-5mm}
\caption{Charged-Higgs branching ratios vs $M_{H^\pm}$, for $\tan\beta=3$ and two light neutral Higgs bosons $h$ and $H$ (125~GeV and 130~GeV) for Type-I/X (left)  and -II/Y (right) with $\sin(\beta-\alpha)=0.7$ (top) and $1$ (bottom). Here, $M_{H^\pm}=M_A$.}
\end{figure}

\begin{figure}[htb]
\label{Fig:feyn-figures-hqq}
\begin{center}
\includegraphics[scale=0.55]{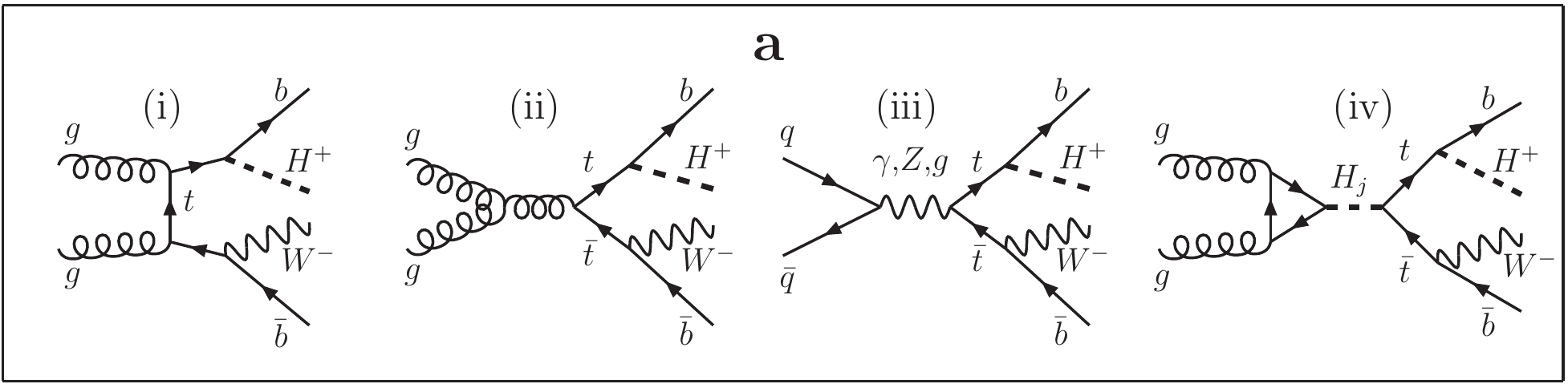}
\includegraphics[scale=0.55]{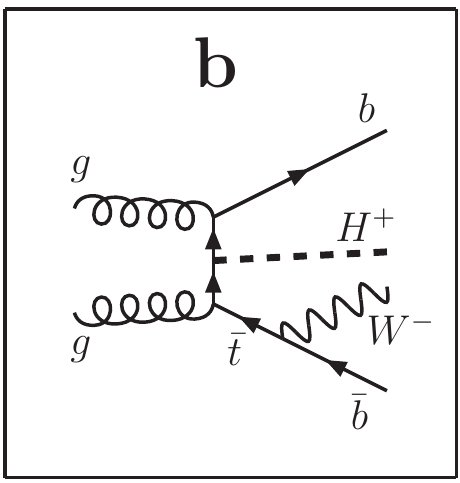}
\end{center}
\vspace*{-5mm}
\caption{Feynman diagrams for the processes $gg,q\bar q\to b\bar b H^+W^-$.  }
\end{figure}

Fig.~4
shows LHC Run 1 (7 and 8 TeV) limits on the model independent production times BR rates for the light and heavy $H^\pm$ range using the $\tau\nu$ decay mode from both ATLAS and CMS while for the $tb$ mode (only applicable to the $M_{H^\pm}>m_t$ case) see
Fig.~5.
Some Run 2 analyses also exist at present, though they do not significantly improve upon the  results shown here.

Furthermore, $H^\pm$ properties can also be accessed  indirectly, through either limits (on any state) or measurements (of the SM-like one, e.g., $H^\pm$ can enter in $h\to\gamma\gamma$ and $Z\gamma$ decays) in the whole Higgs sector. Using {\tt HiggsBounds} \cite{Bechtle:2013wla} and {\tt HiggsSignals}
\cite{Bechtle:2013xfa}, constraints on the [$\cos(\beta-\alpha)$, $\tan\beta$] plane can be drawn for all 2HDMs, as shown in 
Fig.~6.

\begin{figure}[htb]
\label{Fig:lhc-limits}
\vspace*{-2mm}
\begin{center}
\includegraphics[scale=0.09]{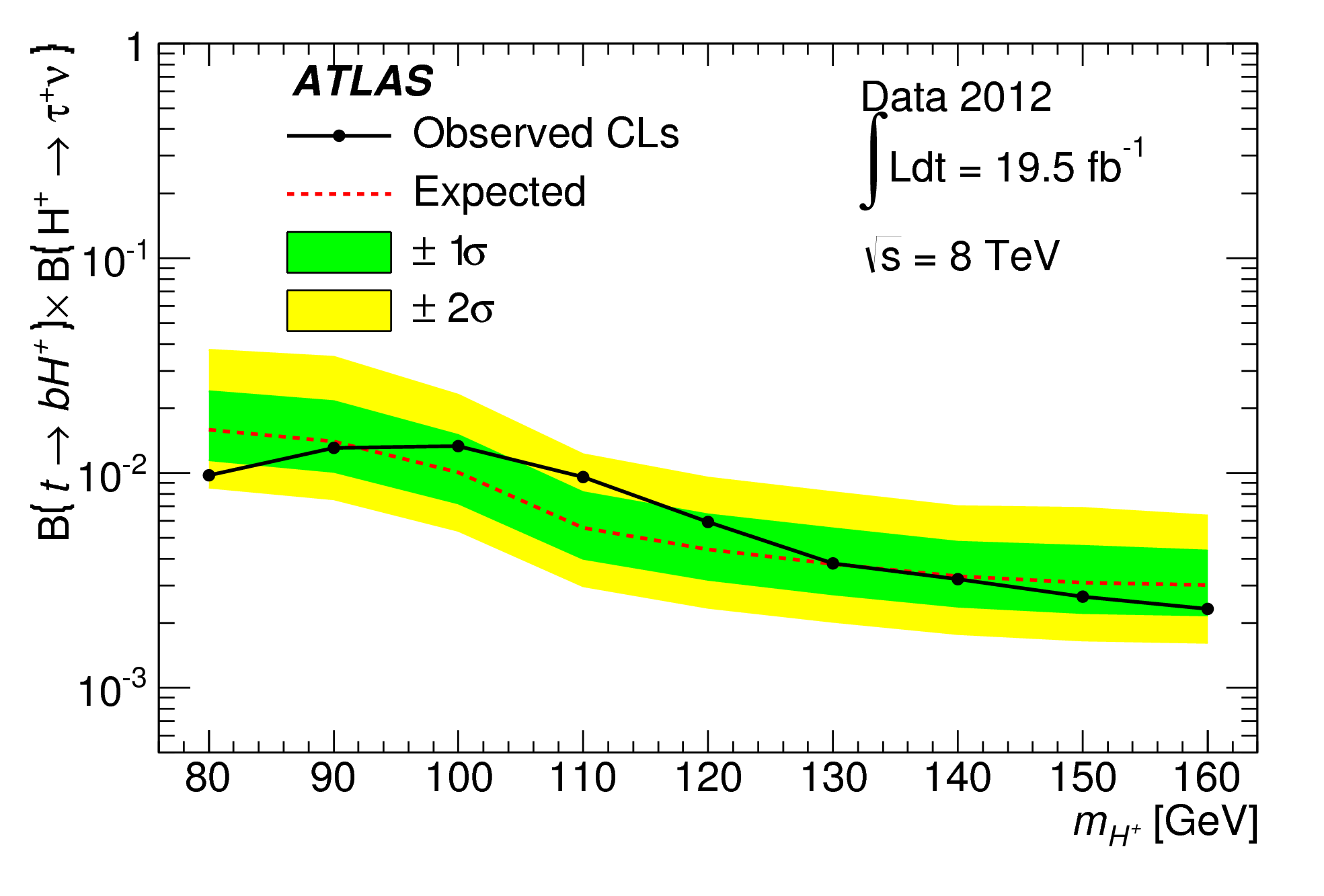}
\includegraphics[scale=0.09]{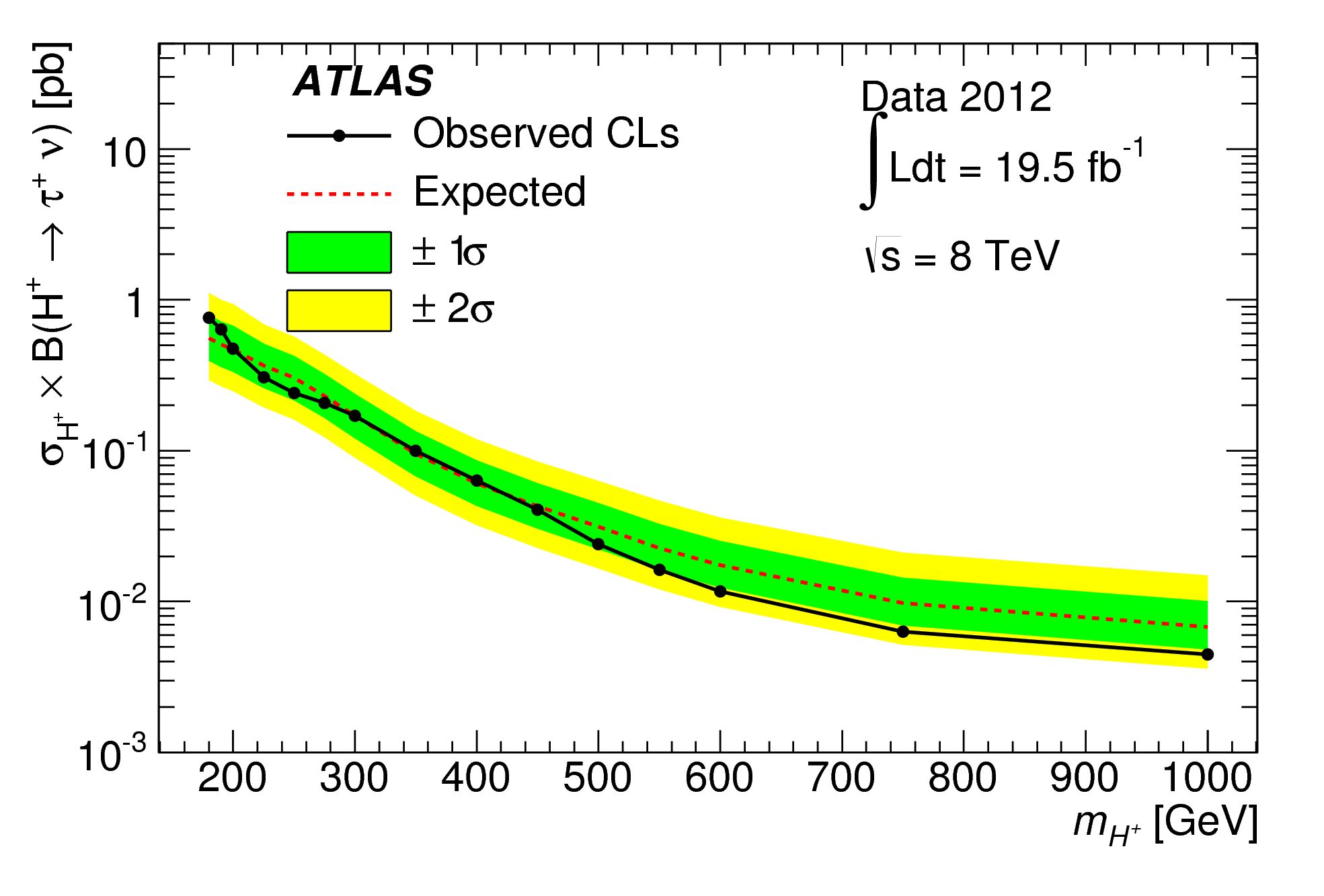}\\
\includegraphics[scale=0.031]{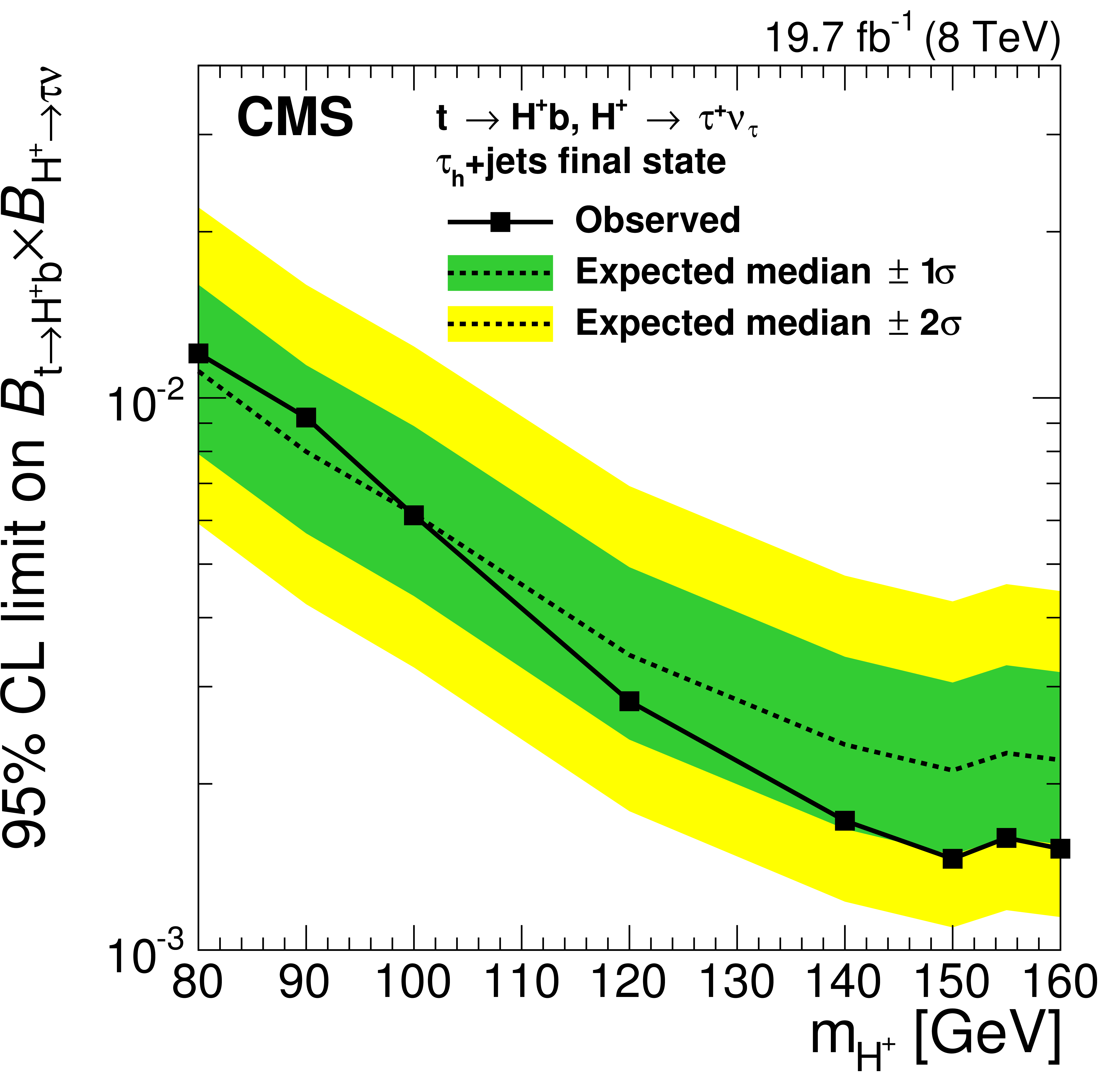}
\includegraphics[scale=0.031]{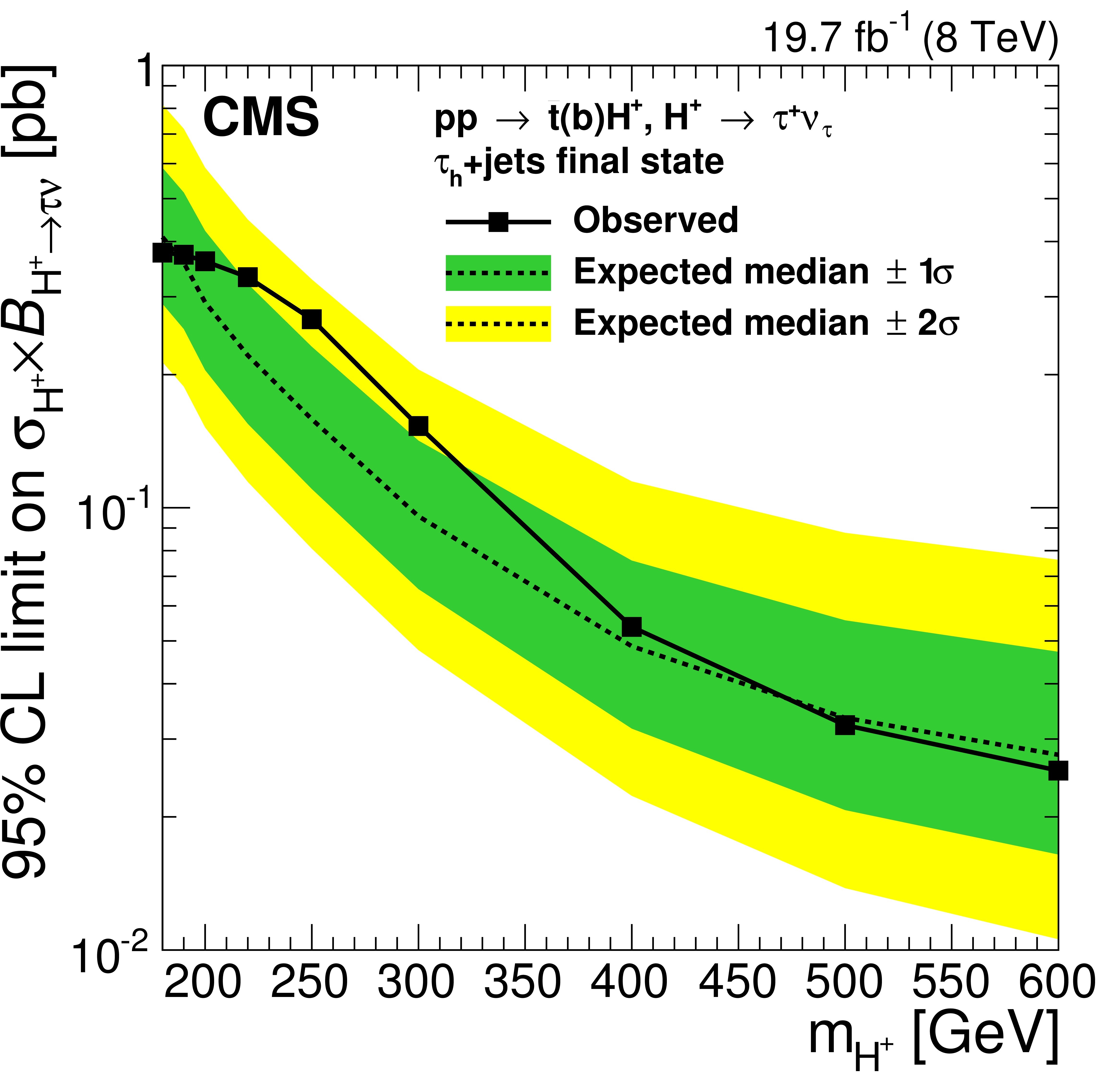}
\end{center}
\vspace*{-7mm}
\caption{ATLAS (top) and CMS (bottom) upper limits on  BR$(t\to H^+b)\times{\rm{BR}}(H^+ \to \tau^+\nu_\tau)$ (left) and $\sigma(pp\to t(b)H^+)\times{\rm{B}R}(H^+ \to \tau^+\nu_\tau)$ (right) rates. [Fig.~7 of \cite{Aad:2014kga} (ATLAS) and Fig.~8 of \cite{Khachatryan:2015qxa} (CMS).]}
\end{figure}

\section{Future LHC Prospects}
\label{sec:future}
While further investigation of the $H^\pm\to\tau\nu$ and $tb$ modes is warranted for Run 2, as intimated, additional  interesting possibilities will be offered
by the $cb$ (in Type-Y) and $W^\pm A$ (in Type-I) channels in the low $M_{H^\pm}$ (and $M_A$) range. The case for exploiting the former (also with a view at measuring $\tan\beta$) was 
already made in \cite{Akeroyd:2012yg} and has now lead (in CMS)  to competitive (with $\tau\nu$) limits (see
Fig.~7) 
while the latter (also
sensitive to $\alpha$) was recently advocated in \cite{Arhrib:2016wpw} (see 
Fig.~8).

\begin{figure}[htb]
\label{Fig:lhc-heavy-only}
\begin{center}
\includegraphics[scale=0.09]{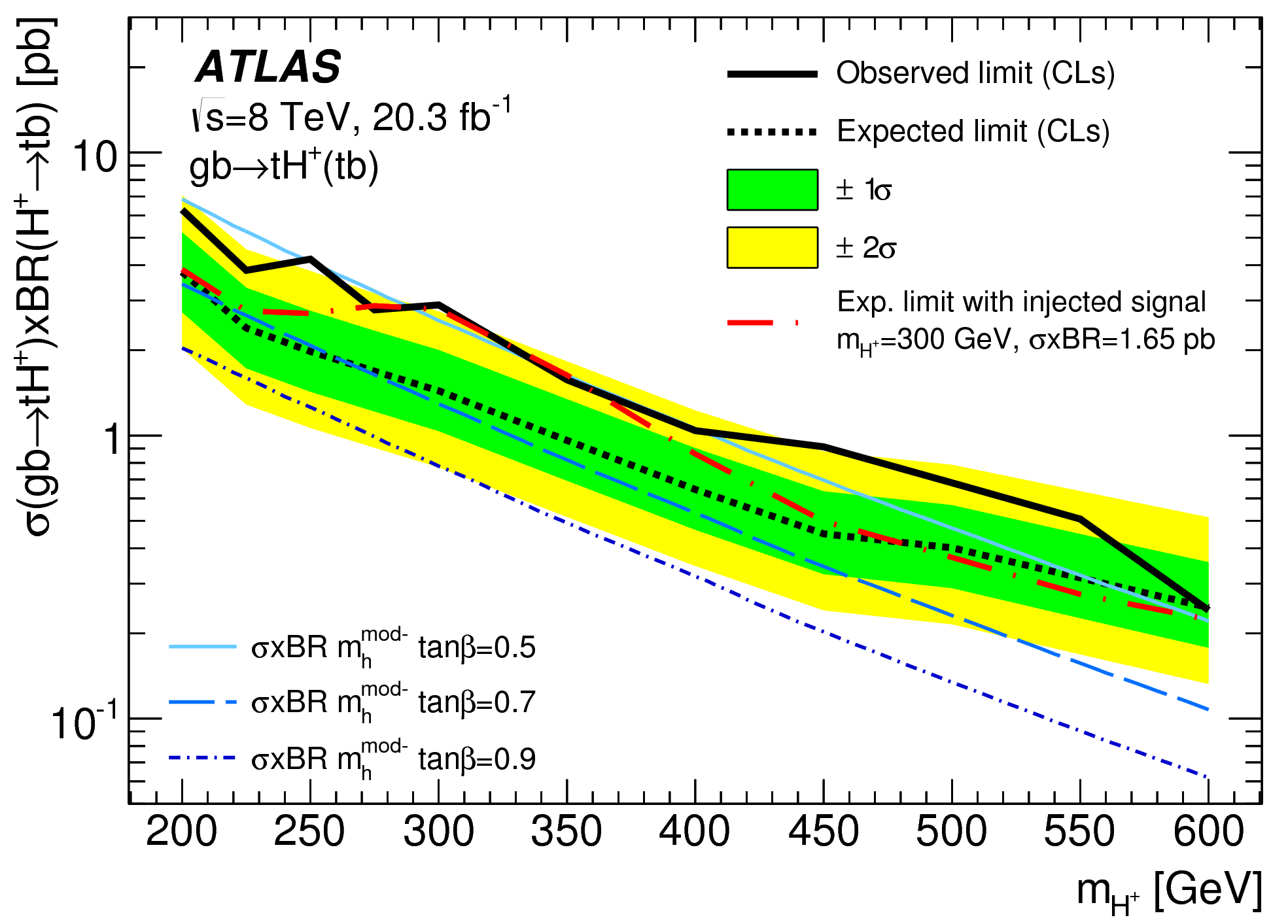}
\includegraphics[scale=0.031]{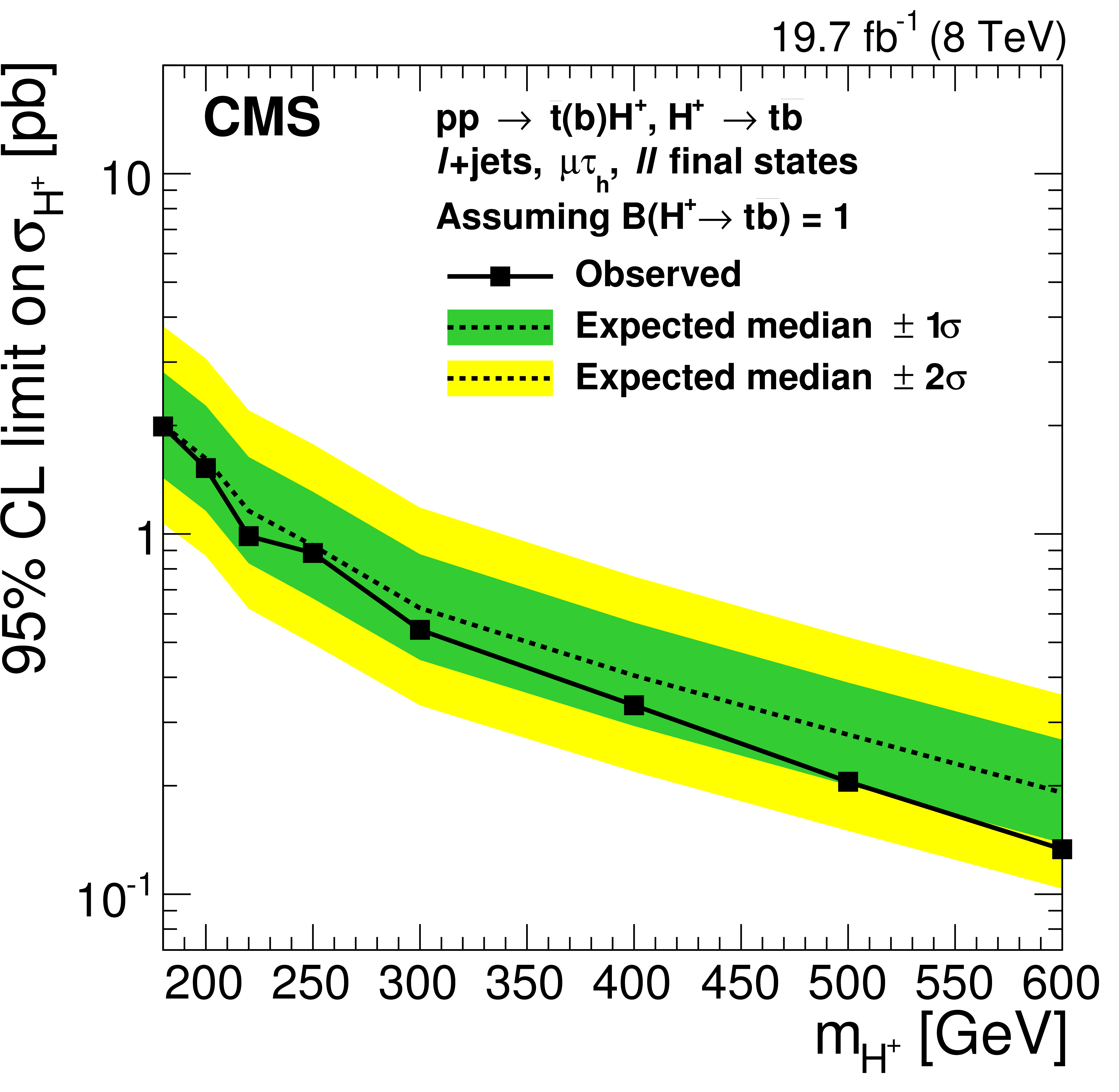}
\end{center}
\vspace*{-7mm}
\caption{ATLAS (left) and CMS (right) upper limits on the  $\sigma(pp\to t(b)H^+)\times{\rm{B}R}(H^+ \to t\bar b)$ rate. 
[Fig.~6 of \cite{Aad:2015typ} (ATLAS) and Fig.~10 of \cite{Khachatryan:2015qxa} (CMS).]}
\end{figure}
\begin{figure}[t]
\begin{center}
\includegraphics[width=0.3\linewidth,angle=270]{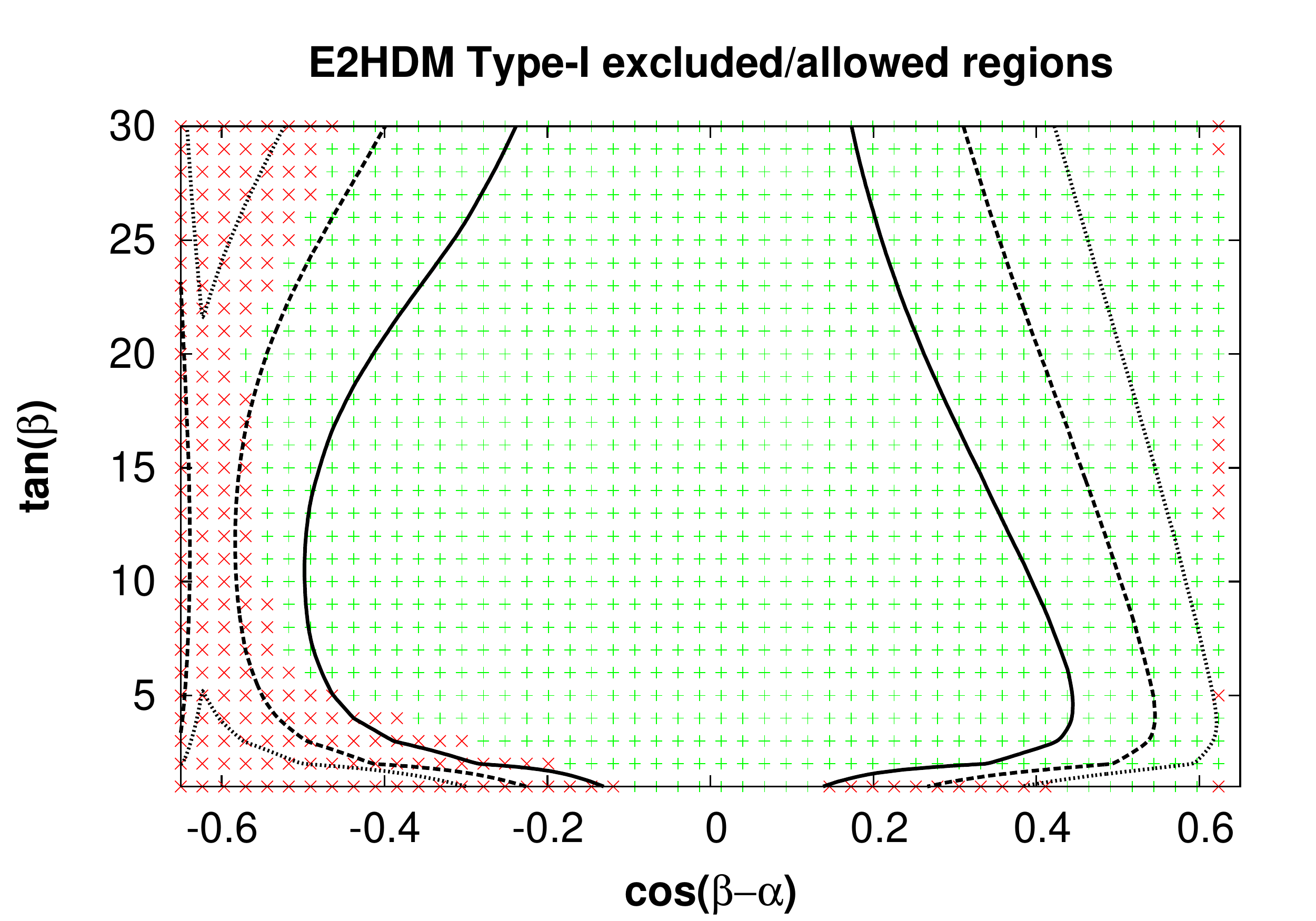}\hspace{4mm}
\includegraphics[width=0.3\linewidth,angle=270]{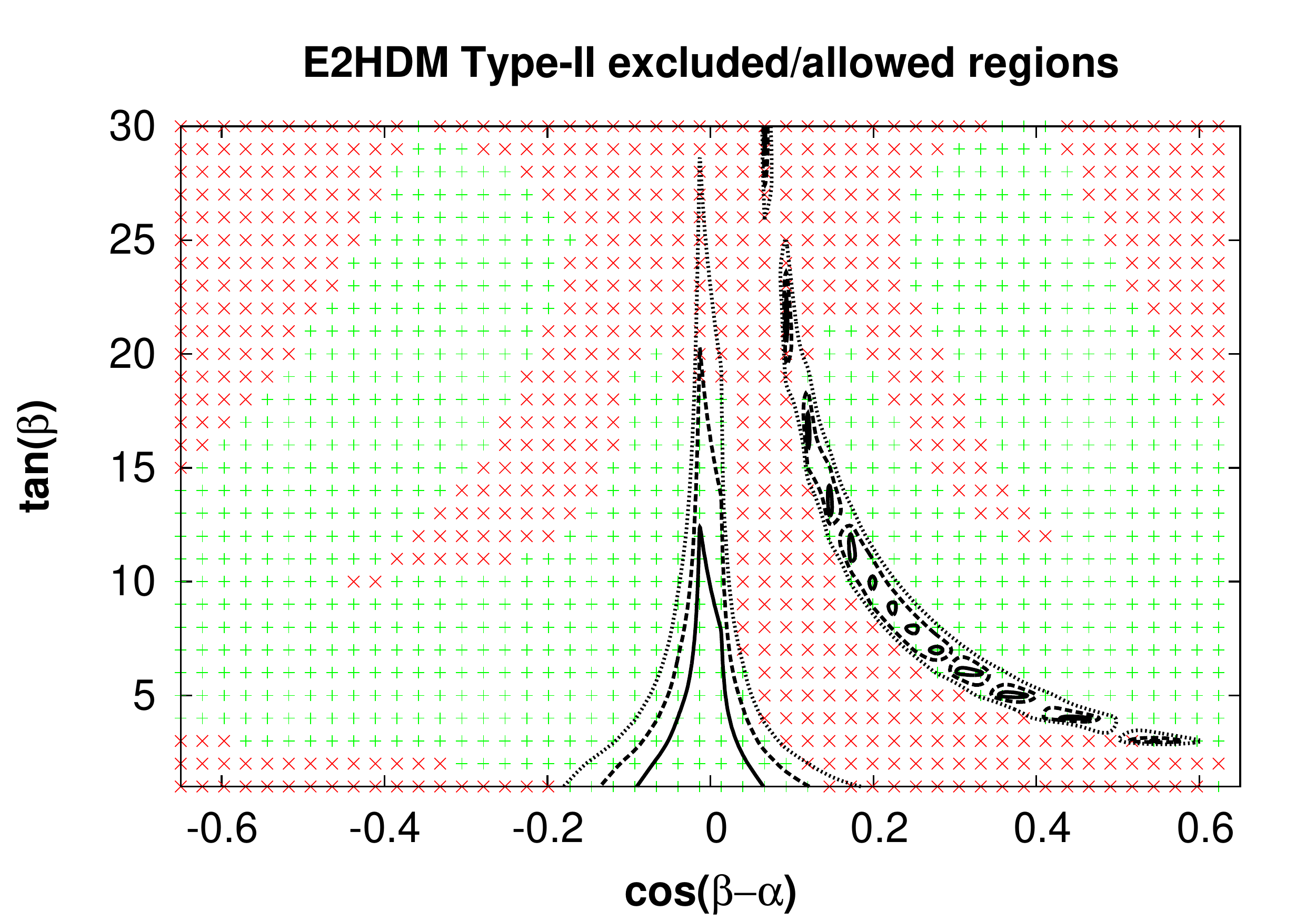}\hspace{3.25mm}\\
\includegraphics[width=0.3\linewidth,angle=270]{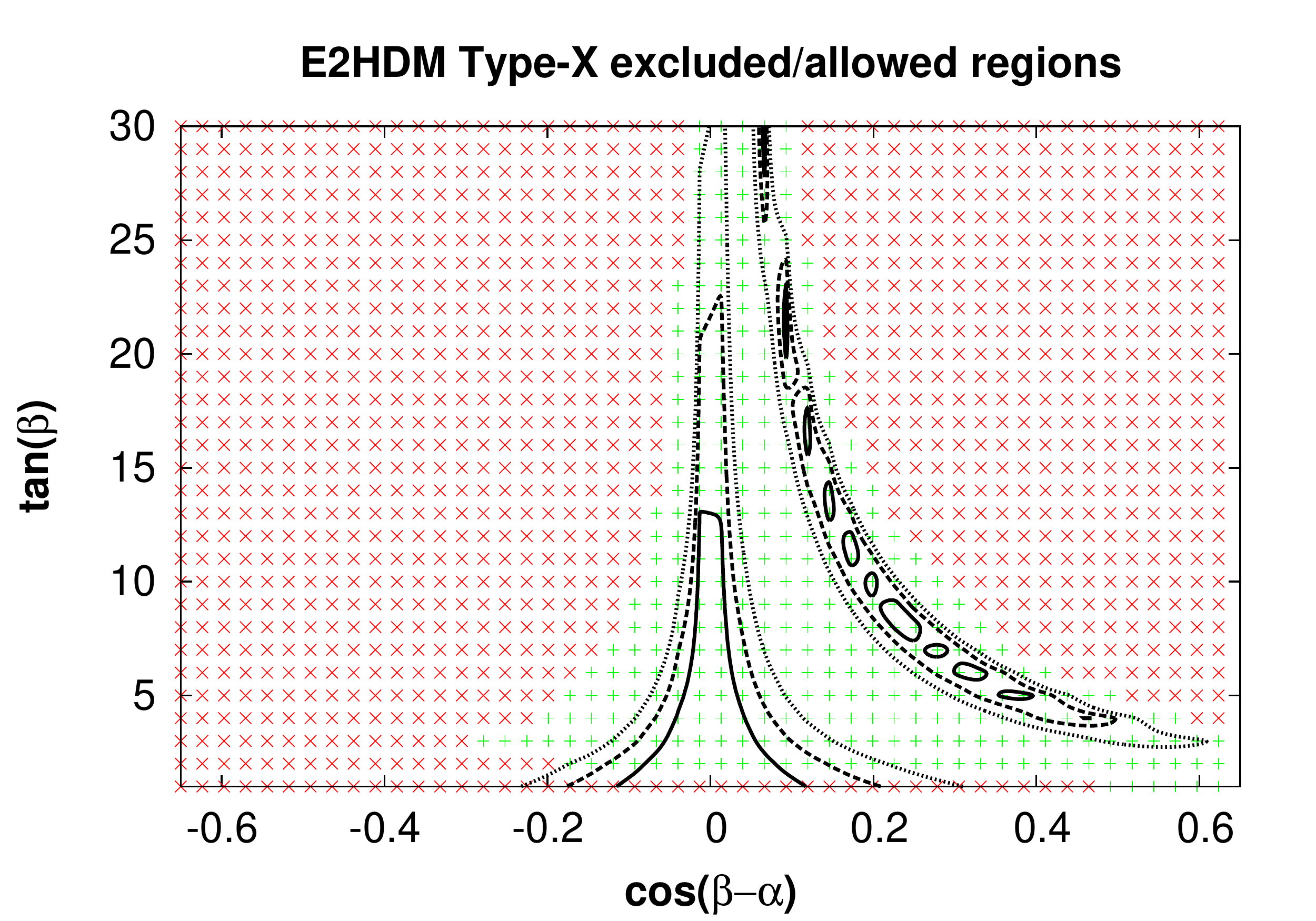}\hspace{3.25mm}
\includegraphics[width=0.3\linewidth,angle=270]{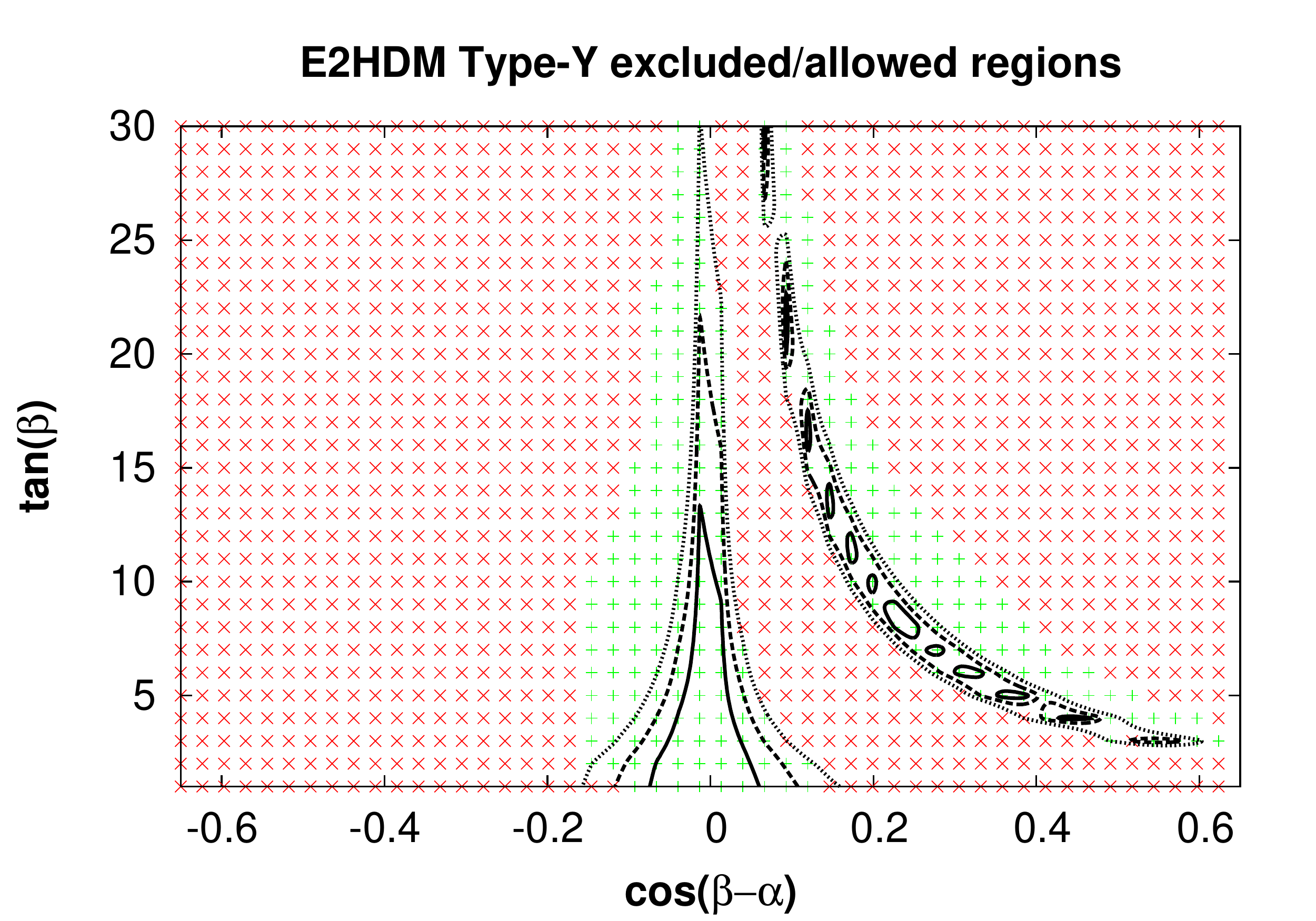}\\ \vspace{5mm}
\caption{Green\textcolor{green}{$\ast$} (Red\textcolor{red}{$ \times $}): allowed (excluded) regions from  LEP, Tevatron and LHC experiments at 95\% CL in all the 2HDMs. 
The solid, dashed and dotted curve display the contour for $ \Delta \chi^{2}=$ 2.30 (68.27\% CL), 6.18 (95.45\% CL) and 11.83 (99.73\% CL),
respectively.  Here,  $m_{h}=125$ GeV and $m_{H}=m_{H^{\pm}}=m_{A}=500$ GeV. 
}
\label{fig:HB-HS}
\end{center}
\end{figure}

\section{Conclusions}
\label{sec:summa}
In summary, several charged Higgs production and decay channels afford the LHC with sensitivity to various Yukawa structures of a 2HDM. Herein, 
current limits from direct $H^\pm$ searches exclude significant portions of parameter space. Yet, for the future, the combination of both established and new (fermionic and bosonic) decays of  (both light and heavy) charged Higgs states will offer one the possibility of both discovery and separation of a specific 2HDM scenario.\\

\noindent{\bf Acknowledgements}~This research  is supported in part through the NExT Institute and by the grant
H2020-MSCA-RISE-2014 no. 645722 (NonMinimalHiggs).

\begin{figure}[!t]
\label{fig:cb}
\begin{center}
\includegraphics[scale=0.40,angle=0]{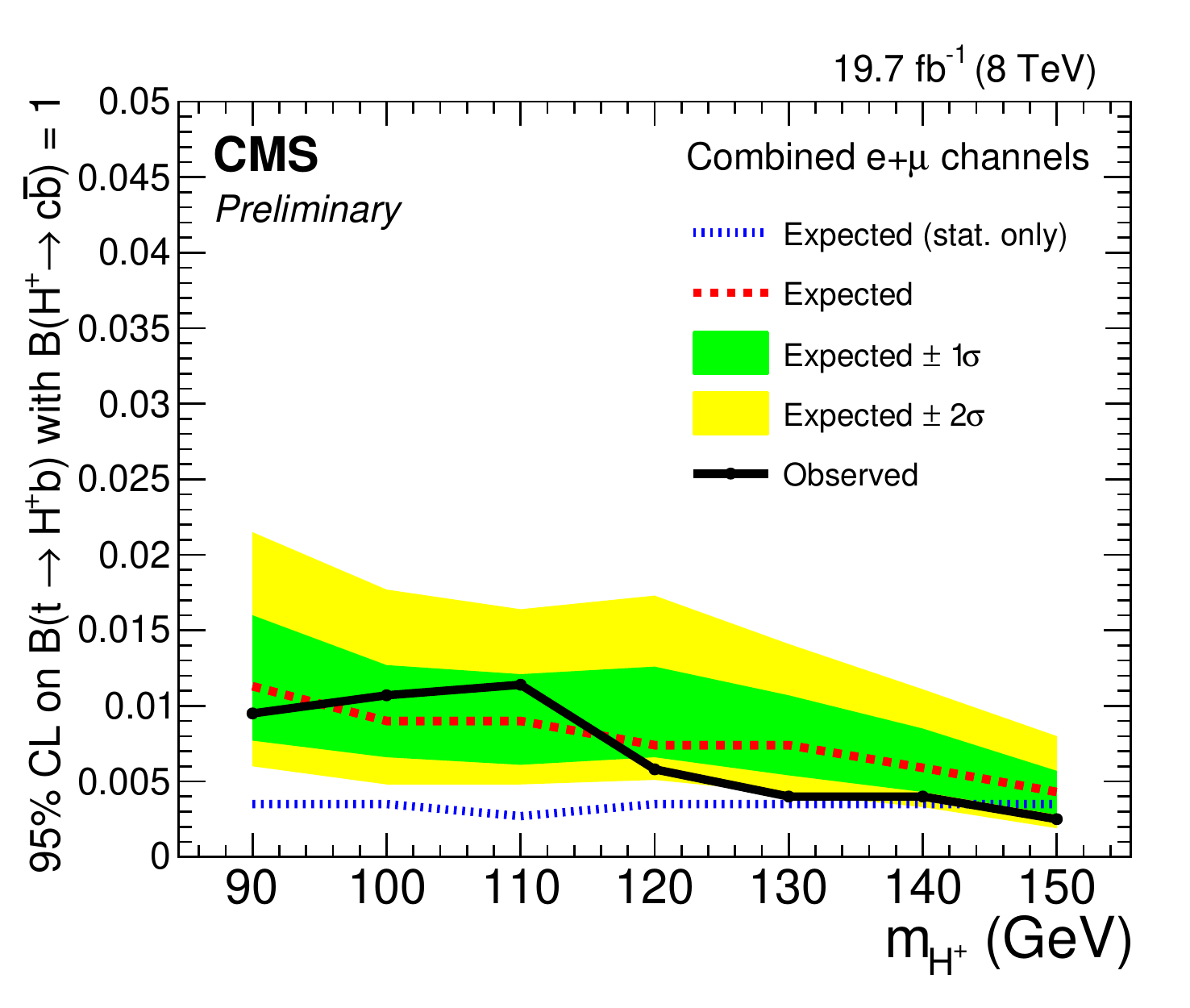}
\end{center}
\vspace*{-7mm}
\caption{CMS upper limit on the  BR$(t\to H^+b)\times{\rm{BR}}(H^+ \to c\bar b)$ rate. [Fig.~14c of \cite{CMS-PAS-HIG-16-030}.]}
\end{figure}

\begin{figure}[!t]
\label{fig:WA}
\begin{center}
\includegraphics[width=0.40\textwidth]{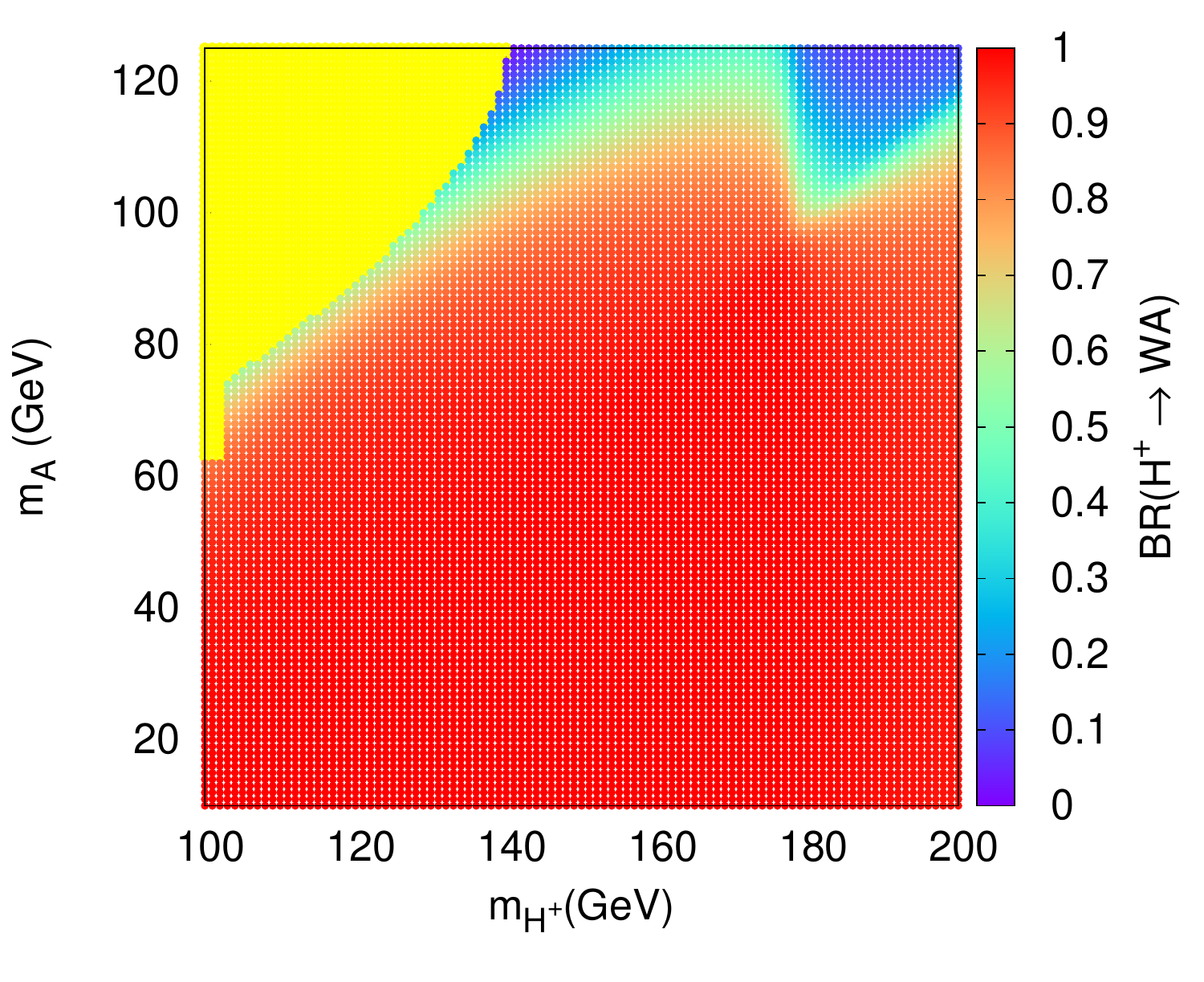}
\end{center}
\vspace*{-9mm}
\caption{BR$(H^\pm \to W^{\pm}A)$   in the 2HDM-I mapped over the $[m_{H^\pm},m_A]$ plane for  $M_h = 125 $ GeV, $\sin(\beta-\alpha) = 1$, $\tan\beta = 5$ 
and $M_{H}  = 300$ GeV. [The yellow region is excluded by LHC data at $95\%$ CL.] }
\end{figure}

\end{document}